\documentclass[nofootinbib,twocolumn,prc,aps]{revtex4-1}

\usepackage{enumerate}
\usepackage{natbib}
\usepackage{amsmath}
\usepackage{amsfonts}
\usepackage{amsthm}
\usepackage{color}
\usepackage{graphicx}
\usepackage{array,multirow,graphicx}
\usepackage{hyperref}
\usepackage{natbib}
\usepackage{color,colortbl}
\usepackage{srcltx}

\def\eps{\varepsilon}
 
\def\fm{\rm fm}
\def\fm2{\rm fm^2}

\newcolumntype{C}[1]{{\centering\arraybackslash}p{#1}}
\definecolor{Gray}{gray}{0.9}

\begin{document}

\title{Multilayer neutron stars with scalar mesons crossing term}

\author{Sebastian Kubis}
\email{skubis@pk.edu.pl}
\author{W\l{}odzimierz~W\'ojcik}
\affiliation{Institute of Physics, Cracow University of Technology,  Podchor\c{a}\.zych 1, 
30-084 Krak\'ow, Poland}
\author{Noemi Zabari}
\affiliation{Henryk Niewodnicza\'nski Institute of Nuclear Physics,\\
Polish Academy of Sciences,\\
Radzikowskiego 152, 31-342 Krak\'ow, Poland}

\begin{abstract}
The standard relativistic mean field model, extended with $\sigma$-$\delta$ meson interaction, is applied to study neutron star (NS) properties.
In the equation of state including such interaction, the phase transition occurs in the core region of NS.
The phase transition leads to multilayer neutron star structure and to the softening of the equation of state (EOS) in a 
moderate range of densities. 
Such an EOS, consequently, makes the neutron star more compact without reducing its maximum mass.
The star properties obtained are in very good agreement with the most recent observational data from the LIGO/Virgo 
gravitational wave detectors and x-ray observations of NICER.  
\end{abstract}

\maketitle

\section{Introduction}

The internal structure of neutron stars (NSs) is one of the most enigmatic subjects of nuclear astrophysics. The latest observation of the neutron-star merger  --  associated with gravitational wave  detection by Advanced LIGO/Virgo -- in August 2017 revealed details on the mass and radius of neutron star determination.
The event is known as 
GW170817 \cite{TheLIGOScientific:2017qsa}. 
The estimations coming from this event have imposed constraints on the equation of state (EOS) of nuclear matter by simultaneous determination of mass and radius of the newly born star. However, precise mass measurements without any information about stellar radius come from rotating neutron stars. 
A neutron star of mass $1.97 \pm 0.04 ~M_\odot$ (PSR J1614-2230) \cite{Demorest:2010bx} and $2.01 \pm 0.04 
~M_\odot$ (PSR J0348-0432) \cite{Antoniadis:2013pzd} has been reported in a binary system with a white dwarf.
In 2018, a pulsar with mass $2.27 \pm 0.17 ~M_\odot$ (PSR J2215-5135) was identified using the Fermi Large Area 
Telescope (Fermi LAT) by spectral line analysis \cite{Linares:2018ppq}.
A feasible way to obtain maximum masses of neutron stars larger than two solar masses is to retain 
befittingly stiff EOS.
However, the typical stiff EOS makes stars less compact 
\cite{Oertel:2016bki} 
which would be in contradiction to the radius estimation coming from GW170817. The typical soft EOS gives a relatively smaller radius but may never reach a mass above 2 $M_\odot$. 
In order to reconcile those two opposite tendencies, we propose a model giving an EOS that is soft at low 
densities (slightly above $n_0$) and stiff at higher densities relevant to the inner part of the NS core.
{This may be achieved by the standard relativistic mean field (RMF) model, which includes only four 
meson fields ($\sigma, \omega, \rho, \delta$) with Yukawa coupling to nucleons, extended with proper meson-meson interaction.
The RMF model with such crossing terms for mesons was proposed in recent decades. For instance, the vector-scalar crossing term -- without $\delta$ meson -- was considered in Ref.~\cite{Mueller:1996pm}. The model with 
 $\delta$ meson and vector-vector crossing term was analyzed in Ref.~\cite{Bunta:2004ej}.}
In this paper, we investigate the neutron star properties in the framework of such a model, which includes the scalar-scalar meson crossing term, which means $\sigma$ to $\delta$ meson coupling. Modification of this type was recently proposed and analyzed in the context of nuclear symmetry energy properties (see Refs.~\cite{Zabari:2019ukk,Zabari:2019clk}). 
The inclusion of such types of interactions leads to supersoft symmetry energy behavior and its slope values that are consistent with most recent studies of nuclear matter properties. In this work, we show results for the neutron stars that utterly conform to the mass and radius constraints coming from not only LIGO/Virgo, but also from the recent results of the NICER observatory.

\section{Model}
The base of the RMF theory gives an adequate Lagrangian, which includes nucleons interacting through the exchange of four mesons: two isoscalar mesons,
 $\sigma$ and $\omega$ and two isovector mesons, $\rho$ and $\delta$.  
In this work, we use the Lagrangian already proposed in Ref.~\cite{Kubis:1997ew}, but it is also enriched with yet another kind of meson-meson interaction, i.e. the
 $\sigma$-$\delta$ meson crossing term \cite{Zabari:2019ukk}.
Thus, the Lagrangian density $\mathcal{L}$ for nucleon and meson fields is the following
\begin{widetext}
\begin{align} \label{lagrangian}
\begin{split}
\mathcal{L}& = 
\frac{1}{2}\left(\partial_\mu\sigma\partial^\mu\sigma-m_\sigma^2\sigma^2\right) -
\frac{1}{4}\left(\partial_\mu\omega_\nu-\partial_\nu\omega_\mu\right)\left(\partial^\mu\omega^\nu-\partial^\nu\omega^\mu\right)+\frac{1}{2}m_\omega^2\omega_\mu\omega^\mu  -
\frac{1}{4}\left(\partial_\mu{\vec{\rho}}_\nu-\partial_\nu{\vec{\rho}}_\mu\right)\left(\partial^\mu{\vec{\rho}}^\nu-\partial^\nu{\vec{\rho}}^\mu\right)+ \frac{1}{2}m_\rho^2{\vec{\rho}}_\mu{\vec{\rho}}^\mu
\\
 &+\frac{1}{2}\partial_\mu\vec{\delta}\partial^\mu\vec{\delta}-\frac{1}{2}m_\delta^2{\vec{\delta}}^2+\bar{\psi}\left(i\partial_\mu\gamma^\mu-m\right)\psi+g_\sigma\sigma\bar{\psi}\psi -g_\omega\omega_\mu\bar{\psi}\gamma^\mu\psi - \frac{1}{2}g_{\rho}\vec{\rho}_{\mu}\bar{\psi}\gamma^\mu\vec{\tau}\psi+g_\delta\vec{\delta}\bar{\psi}\vec{\tau}\psi-U(\sigma) +{\cal L}_{\sigma \delta}, 
\end{split}
\end{align}
\end{widetext}
where $m$ is the nucleon mass, $m_i$ are meson masses and $g_i$ are coupling constants for nucleons to
the corresponding mesons: $i\!=\!\sigma, \omega, \rho$ and $\delta$ 
respectively. The $U(\sigma)$ is the self-interaction of the $\sigma$ meson.
Due to the inclusion of
 $\sigma$-$\delta$ meson interaction, the corresponding part $ {\cal L}_{\sigma \delta}$ is added to the Lagrangian. 
It represents the $\sigma\textrm{-}\delta$ interaction and takes the form 
\begin{equation}
  {\cal L}_{\sigma \delta} = -\tilde{g}_{\alpha} \sigma^{\alpha} \vec{\delta}^2 
  \label{cross}
\end{equation}
where the $\alpha$ exponent distinguishes between two types of meson-meson interactions: $\alpha=1$ for linear and $\alpha=2$ for 
quadratic ones,
where $\delta$ denotes the mean value of the third component of the  $\vec{\delta} $ field. 
In the framework of RMF, for the considered model, there are seven free model parameters represented by the 
following coupling constants: four constants for nucleon-meson couplings $C_i = g_i/m_i ~ i=\sigma, \omega, \rho, \delta$,
 two constants $b, c$ for $\sigma$ meson self-interaction (in potential $U(\sigma)$), and one for the $\sigma$-$\delta$ crossing term
  $\tilde{g}_\alpha$. 
In the model parametrization, it is more convenient to use 
$g_\alpha = \displaystyle\frac{\tilde{g}_\alpha}{4 g_\sigma^\alpha g_\delta^2}$ instead of $\tilde{g}_\alpha$.
Then, the additional contribution to the total energy density of matter coming from the $\sigma$-$\delta$ crossing term
is
\begin{equation}
\epsilon_{\sigma \delta} = g_\alpha \sigma^\alpha \delta^2.
\label{cross-term}
\end{equation}
 For more details, we refer to Ref.~\cite{Zabari:2019ukk}.
 The isoscalar sector consists of $C_\sigma$, $C_\omega$, $b$, and $c$ coupling constants, which values are attained 
 to reproduce the symmetric nuclear matter. In our calculations, we adopt the saturation point density 
  $n_0 = 0.16 ~\rm fm^{-3}$, the binding energy $B=-16 ~\rm  MeV$ and the compressibility $K(n_0)=230 ~\rm MeV$. 
 
 {The presence of scalar mesons $\sigma $ and $\delta$ modifies the effective nucleon mass in the following way: 
  \begin{align}
 \label{meff-proton}
m_p^\ast=m-g_\sigma\bar{\sigma}-g_\delta{\bar{\delta}}_3 ,\\
\label{meff-neutron}
m_n^\ast=m-g_\sigma\bar{\sigma}+g_\delta{\bar{\delta}}_3 .
\end{align}
For symmetric matter, they are equal to each other $m^* \equiv  m_p^*=m_n^*$ as the $\delta$ mean field is zero.
Experiments give some estimation of the effective nucleon mass at the saturation point $m^*(n_0)$. The commonly accepted value, coming from 
isoscalar giant quadrupole resonances, is $m^*(n_0)/m = 0.8 \pm 0.1$,  \cite{Li:2018lpy}. Our model predicts the effective nucleon mass at $n_0$ to be 0.73, which is still a reasonable value.
One must remember that the effective masses expressed by Eqs.~(\ref{meff-proton} and \ref{meff-neutron}) are the Dirac masses and that they differ slightly from the other definitions of effective masses (nonrelativistic ones) usually used to fit the experimental data.}
 
 The stiffness of the EOS is mainly controlled by the constant $C_\sigma$. Here we adopt it to be $11 ~\fm2$ 
 which gives sufficiently stiff EOS that is compatible with observations of PSR J2215-5135 and PSR J0740-6620. 
 
\begin{figure}[t!]
\begin{center}
\includegraphics[width=\columnwidth]{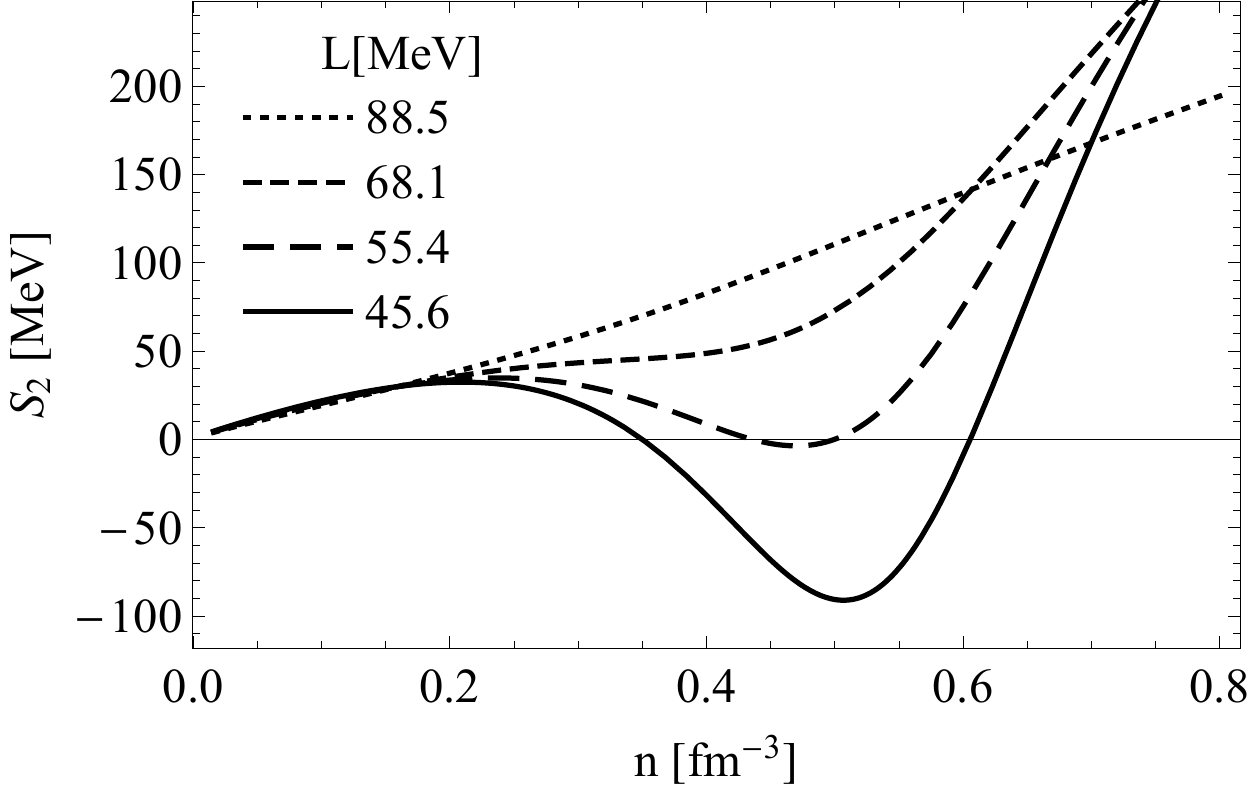}
\caption{Symmetry energy as a function of density for models with different $L$.}
\label{Esymplot}
\end{center}
\end{figure}

 { The setting of couplings in the isovector sector,  $C_\rho$, $C_\delta$, and $g_\alpha$ requires deeper discussion. 
 These three constants are fitted to two observables: the symmetry energy 
 at saturation $S_2(n_0) = 30$ MeV and the symmetry energy slope
 $L= 3n_0 \frac{dS_2}{dn}|_{n_0}$. Thus, one of them remains free. 
 Recent experimental data suggest a rather low slope in the range between
 40 and 80 MeV. Such low values are obtained thanks to the introduction of the $\sigma$-$\delta$ crossing term Eq. (\ref{cross}).
 Without the crossing term, the value of slope $L$ is greater than 90 MeV, which was already noted in Ref.~
 \cite{Liu:2001iz}. The 
 required lower values of $L$ may be achieved if and only if ${g}_\alpha$ takes negative values. 
  The value of $g_\alpha$ cannot be too large; 
 typically, it should be at the order of $10^{-3}$ (for both linear and quadratic types of coupling); otherwise one obtains 
 the effective mass with pathological behavior (discontinuities or negative mass). In this work, we focus on the quadratic 
 model ($\alpha=2$). Conclusively, we adopt $g_2=-0.004$. Such 
 setting of $\sigma$-$\delta$ coupling allows for control of the slope $L$ in the range between 40 and 80 MeV by changing 
 the other constant from the isovector sector, i.e. 
 $C_\delta$. To summarize, we obtained a family of models with different symmetry energy slopes, controlled by the $C_\delta$  coupling constant, maneuvering its values from 1 to 3.8 $\fm2$. 
 
The symmetry energy as a function of density for these models is shown in Fig.~\ref{Esymplot}. The value of the slope $L$ at the saturation point has dramatic consequences for symmetry energy behavior at higher densities. The symmetry energy bends strongly and may even become negative. Such negative values of $S_2$ lead to instability of charge fluctuations 
\cite{Kubis:2006kb} and finally result in the phase separation. The phase transition and the construction of the two-phase system are presented in detail in the next section.}

 \section{Phase separation}
 
\begin{figure}[t!]
\begin{center}
\includegraphics[width=\columnwidth]{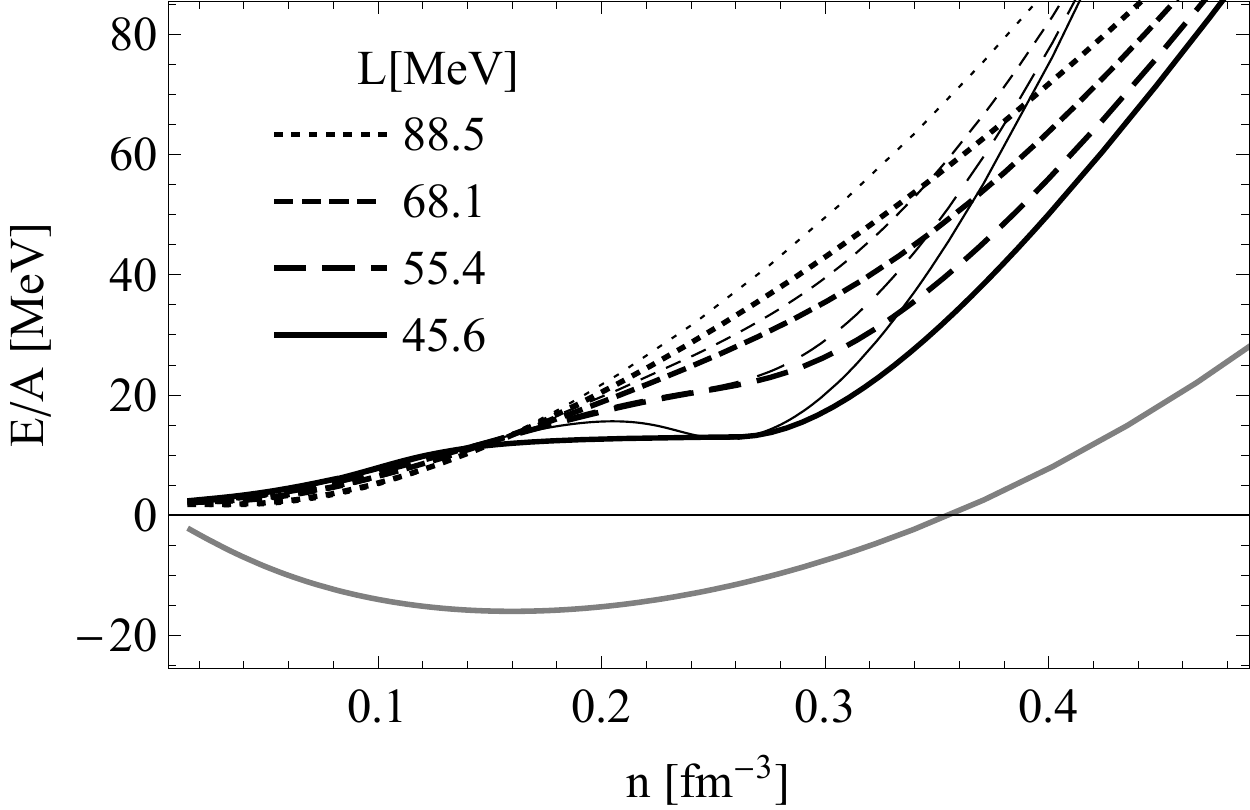}
\caption{Energy per baryon in dense matter for different models. Thick and thin black lines correspond to matter in $\beta$-equilibrium and pure neutron matter, respectively. The gray line represents the symmetric matter.}
\label{enparplot}
\end{center}
\end{figure}

\begin{figure}[t!]
\begin{center}
\includegraphics[width=\columnwidth]{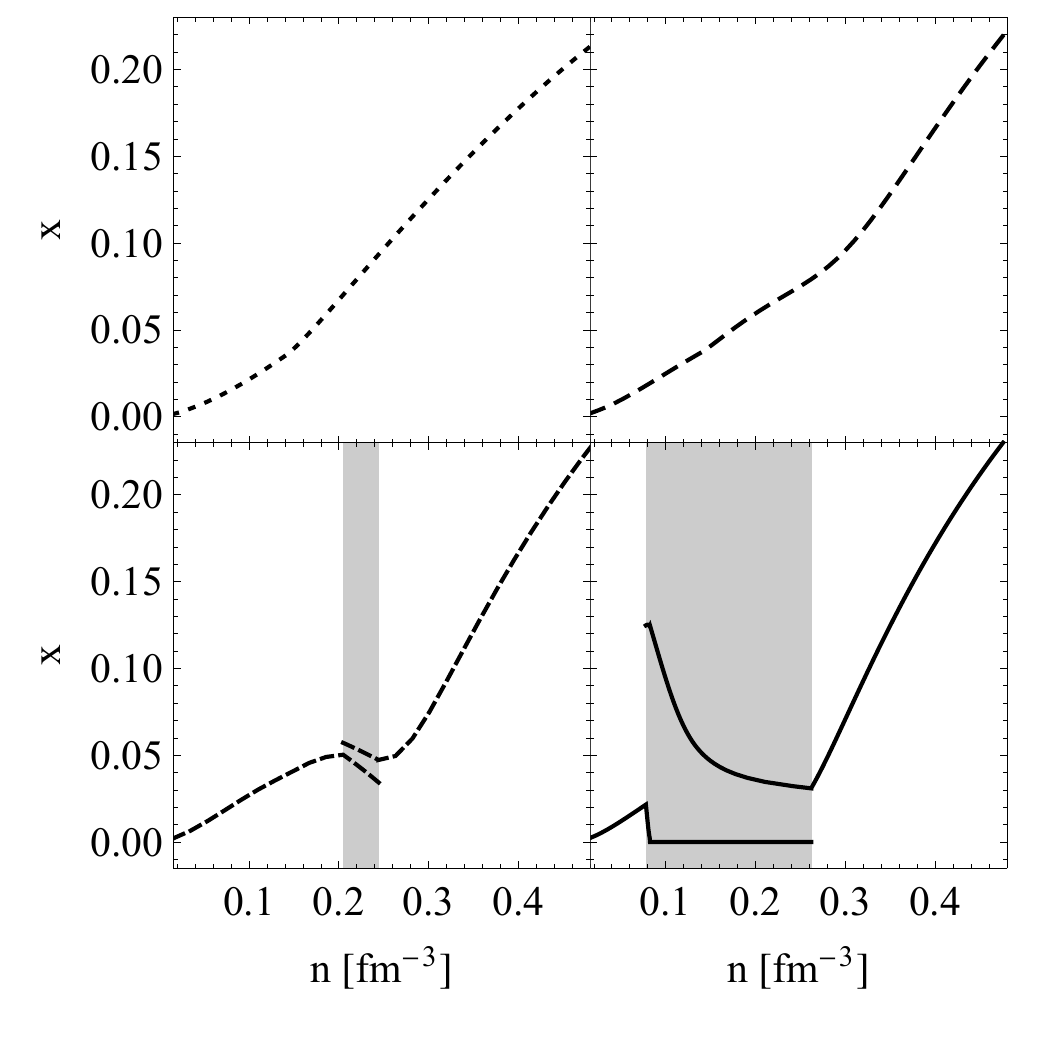}
\caption{The proton fraction in different models (the types of lines correspond to the different $L$). The gray region indicates the system with two phases with two different proton fractions.}
\label{xplots}
\end{center}
\end{figure}

As mentioned above, we use the quadratic model, as it was found that 
 in this model family with quadratic coupling, the phase transition at the low-density region emerges and leads to 
 the softening of the EOS, which is a characteristic property of phase heterogeneity.
 {The EOS softening is already signaled by the flattening of the density dependence of the energy per baryon, 
 which is seen in Fig.~\ref{enparplot}}.
  For a given nuclear model, we build the EOS of nuclear matter in $\beta$-equilibrium. The chemical potentials of each 
 particle species satisfy the $\beta$-equilibrium equality
\begin{equation}
\mu_e = \mu_\mu = \mu_n -\mu_p ,
\end{equation}
where $n$, $p$, $e$, and $\mu$ stand for neutrons, protons, electrons and muons respectively. 
 In order to control the slope value $L$, we manipulate the constant
 $C_\delta$ while $g_\alpha$  is fixed as it is shown in Table~\ref{tab-ns}
- the higher the $C_\delta$ the lower value of the slope. 
\begin{figure}[t!]
\begin{center}
\includegraphics[width=\columnwidth]{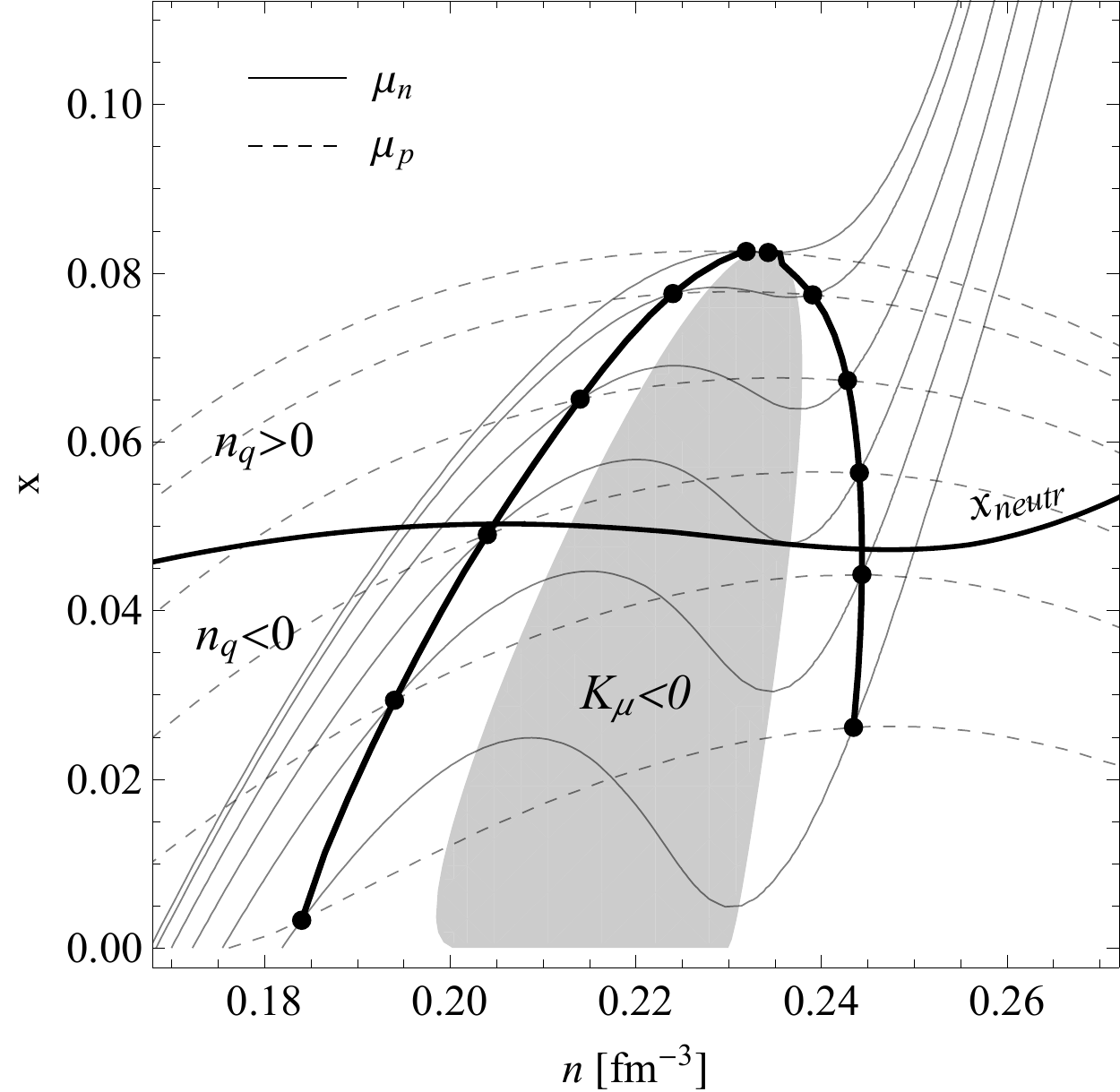}
\caption{
Phase coexistence diagram for the model with quadratic interaction and $C_\delta^2 = 3.5~ \fm2$.
 The thick line is the binodal curve with marked points being in chemical and mechanical equilibrium. The gray area denotes unstable phases -- the spinodal region.}
\label{pt35}
\end{center}
\end{figure}
 Analysis reveals the occurrence of phase transition for high enough $C_\delta$. 
Indeed, such models appear to have negative values of incompressibility $K_\mu$, which signals a split into a two-phase system for the densities where $K_\mu<0$ emerges. The quantity
\begin{equation}
  K_\mu = \left( \frac{\partial P}{\partial n} \right) _\mu
\end{equation}
represents incompressibility under constant chemical potential of charge ($\mu \equiv \mu_e =\mu_\mu$). 
It is the proper quantity to determine the region where the system ceases to be stable against charge 
fluctuations and must split into two phases with opposite charge densities 
  \cite{Kubis:2006kb}.
 The coexistence of those two phases occurs when their pressures and chemical potentials for particles being present in both phases are equal to each other:
\begin{eqnarray}
  P(n^I,x^I) &= &P(n^{II},x^{II}) \nonumber \\
  \mu_i(n^I,x^I)& =& \mu_i(n^{II},x^{II})~~~,~i=n,p,e.
 \label{gibbs}
\end{eqnarray}
These equations, called Gibbs conditions, ensure the mechanical and chemical equilibrium between phases. They lead to what is called Gibbs construction, which is shown on the proton fraction versus density diagram,
Fig.~\ref{pt35}.
 The spinodal region, marked with a gray area, represents the points where incompressibility is negative.  
 No stable phase arises therein. The Gibbs conditions Eqs.~(\ref{gibbs})
 are fulfilled for pairs of points for which $\mu_p$ and $\mu_n$-contours intersect 
on the $n\!-\!x$ space. 
These points form the curve called the binodal line.  
In Fig.\ref{pt35}, the line labeled $x_{neutr}$ represents the locally neutral phase.
 Above that line, all phases are positive ($n_q>0$); below it, they are negative ($n_q<0$). 
Global charge neutrality assures that only points for phases with opposite charge may coexist in the real system.  
Those two phases have different proton abundances, $x^I$ and $x^{II}$ as shown in Fig.~\ref{xplots}. Phase I always has a smaller proton fraction and is more dilute $n^I < n^{II}$. With increasing mean density of the system, it disappears in favor of the phase II, and its proton fraction becomes very low. For the model with 
the lowest $L$, the phase I at some density becomes the pure neutron matter, $x^I =0$, which is clearly visible in the fourth panel of Fig.~\ref{xplots}.
The global neutrality of the system determines the volume proportion between phases expressed
by the equation
\begin{equation}
  w(n_p^{I}-n_e^{I}-n_\mu^{I}) + (1-w)(n_p^{II}-n_e^{II}-n_\mu^{II}) =0 ,
  \label{charge-neutr}
\end{equation}  
where $w$ is the volume fraction occupied by the  phase I: $w = V^{I}/(V^{I}+V^{II})$.
The Gibbs conditions Eq. (\ref{gibbs}) and global charge neutrality Eq. (\ref{charge-neutr})
allow for finding the region of phase space where two separated phases occur and uniquely
determine the equation of state, i.e. the pressure-versus-energy-density relation, $P(\eps)$. 
 It is worth noting that the phase separation appears for the quadratic interaction model with the most plausible 
 value of slope $L$. For the presented model with $C_\delta^2=3.5~\rm \fm2$ and $\alpha=2$, the slope is $L=55.4$ MeV. 
 However, even lower values might be obtained for higher $C_\delta^2$ .
\begin{figure}[t!]
\begin{center}
\includegraphics[width=\columnwidth]{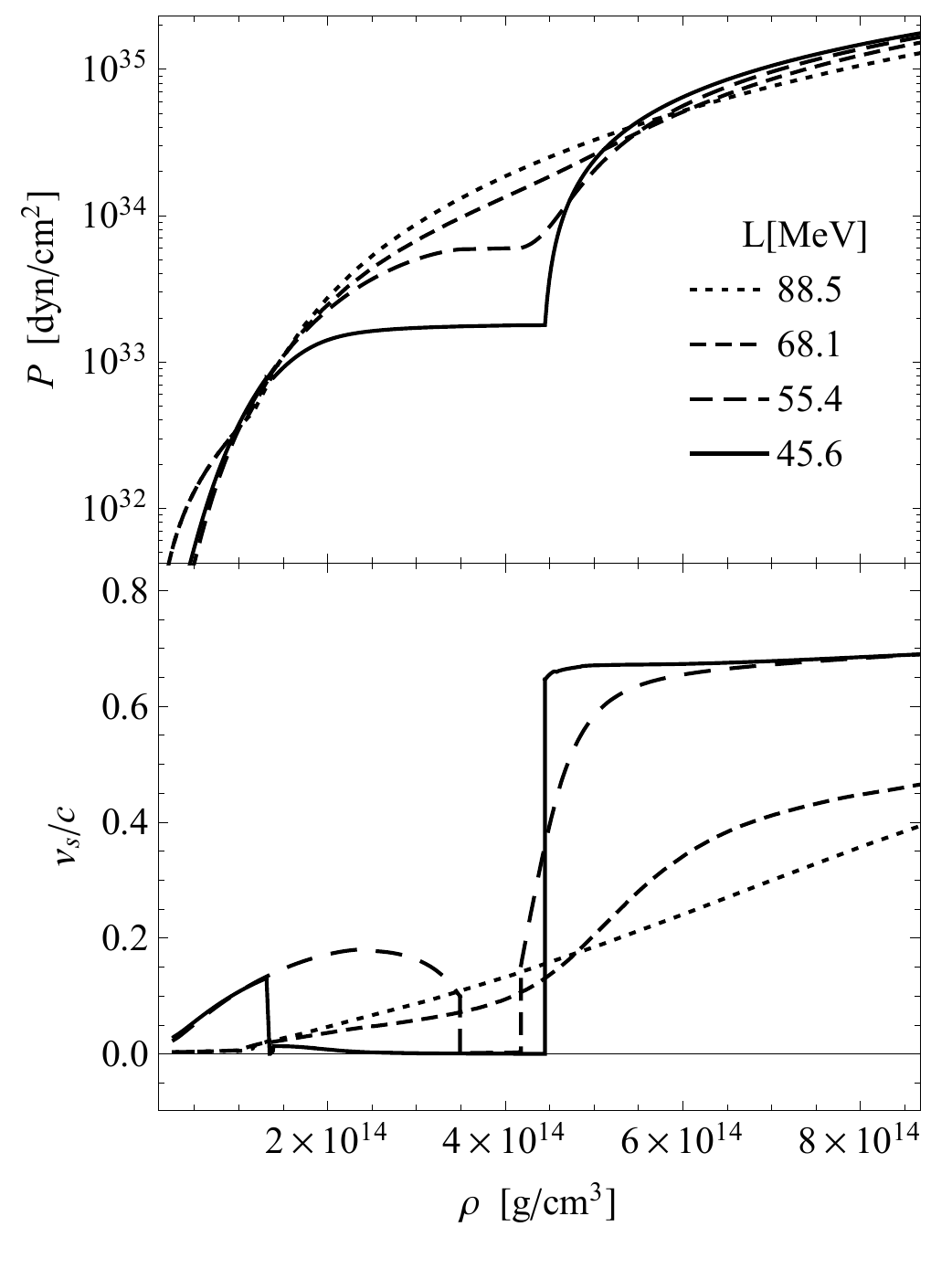}
\caption{The EOS and speed of sound for four different $C_\delta^2$ couplings. Corresponding symmetry energy slopes $L$ are
 given in the legend.}
\label{eossound}
\end{center}
\end{figure} 
The phase separation usually leads to the softening of the EOS in comparison to the one-phase system.
This phenomenon is clearly visible in Fig. \ref{eossound} (upper panel), which shows a set of EOSs for different nuclear models. The phase separation dramatically changes the slope of the $P(\rho)$ relation for the coupling 
$C_\delta^2$ exceeding $3.0~ \fm2$, which corresponds to the slope $L$ being smaller than 68 MeV.
The derivative of the pressure with respect to density determines the sound velocity in the matter
\begin{equation}
 v_s = \sqrt{\frac{\partial P}{\partial \rho}}  ~.
\end{equation}
In the lower panel of Fig. \ref{eossound}, the sound speed is shown in a moderate range of densities. 
 The sudden decrease of $v_s$ corresponds to the phase transition region within which the speed of sound drops to around $10^{-3} c$.

\section{MULTILAYER NEUTRON STARS}

The set of EOSs was constructed for the quadratic model for the $C_\delta^2$ coupling within 1.0 to 3.8 $\fm2$. 
The corresponding values of the symmetry energy slope $L$ are from 88.5 to 45.6 MeV. 
The proposed nuclear RMF model does not cover the region of very low densities, where the neutron star
 crust is formed. For the crust, the Sly4 model was used \cite{Chabanat:1997un, 
 Douchin:2001sv}.
The joining of crustal EOS and the EOS for the core is carried out at the point where their
 pressures and densities are equal. The specific values of crust-core transition densities $n_{c-c}$ for different $L$ are 
 shown in Table~\ref{tab-ns} 

\begin{figure}[b!]
\begin{center}
\includegraphics[width=\columnwidth]{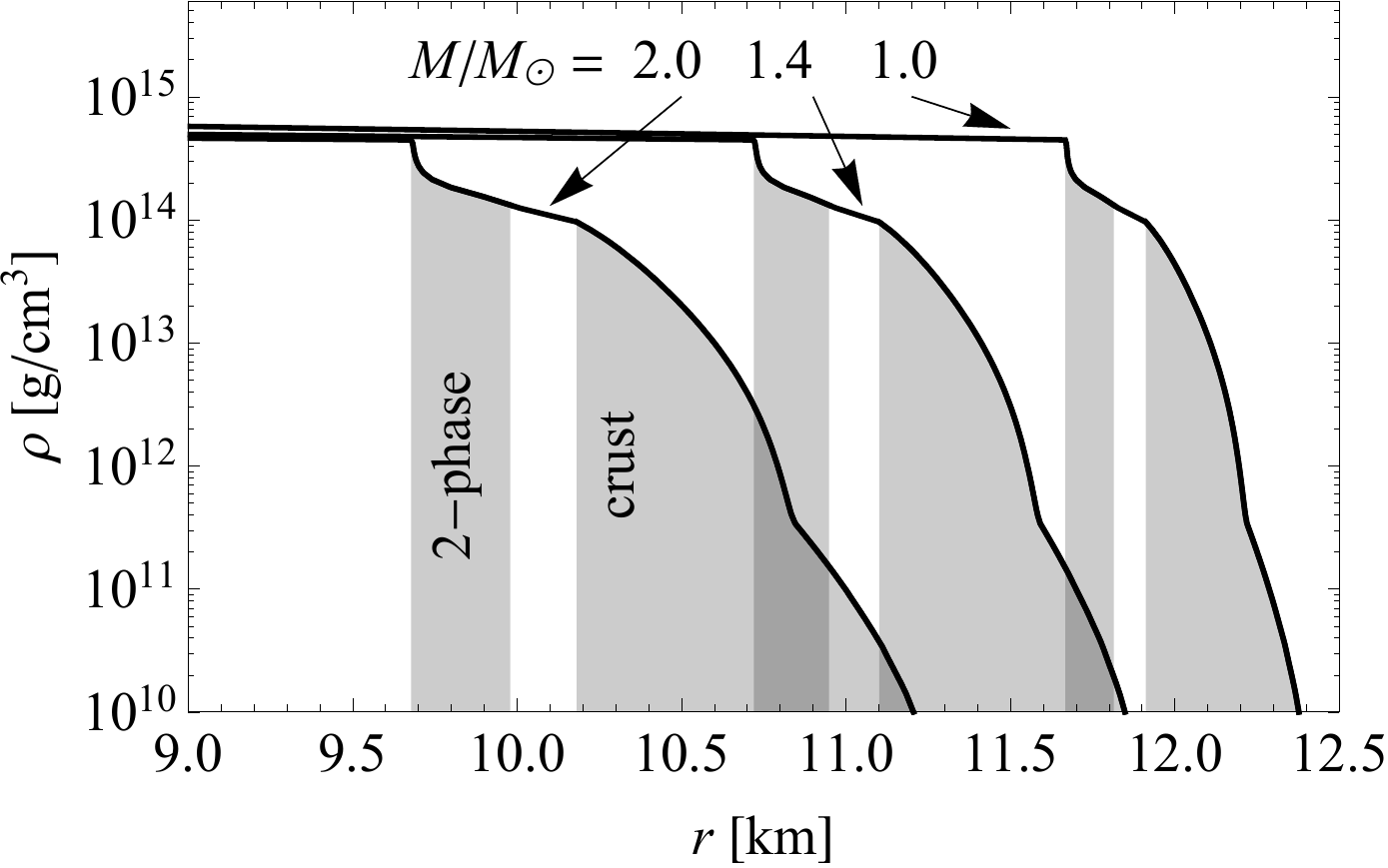}
\caption{Density profiles for different stellar masses for the model with $L=45.6$~MeV.
 The gray color indicates the solid-like structures.}
\label{rho-profile}
\end{center}
\end{figure}

The structure of a nonrotating NS is obtained by implementing given EOSs into Tolman-Oppenheimer-Volkoff
(TOV) equations
\begin{equation}
\begin{split}
&P'(r) = - G\left( \epsilon(r) +P(r) \right) \frac{m(r)+4 \pi r^3 P(r)}{r (r-2G m(r))},\\
&m'(r) = 4 \pi r^2 \epsilon(r),
\end{split}
\end{equation}
where $P$ and $\eps$ are pressure and energy density, and $m(r)$ is the gravitational mass (in units of energy)
confined inside a sphere with radial coordinate $r$.
The TOV equations reveal the nontrivial structure of a star for the EOS with phase transition. The matter in the region of 
phase separation resembles the structure of inner crust -- clusters with high-proton fractions immersed in low-proton 
environments or pure neutron matter, with solid-state properties. Such a layer is separated from the actual star crust by 
a thin layer of homogeneous liquid matter. To conclude, the star includes the four different layers: two solid-like and two 
liquid. Figure~\ref{rho-profile} shows the subsequent layers with different properties.
Such a multilayer structure of a neutron star requires further analysis in the context of rotational and vibrational properties. The solid-like internal layer should somehow manifest in the pulsar glitching or precession.

\begin{figure}[b!]
\begin{center}%
\includegraphics[width=\columnwidth]{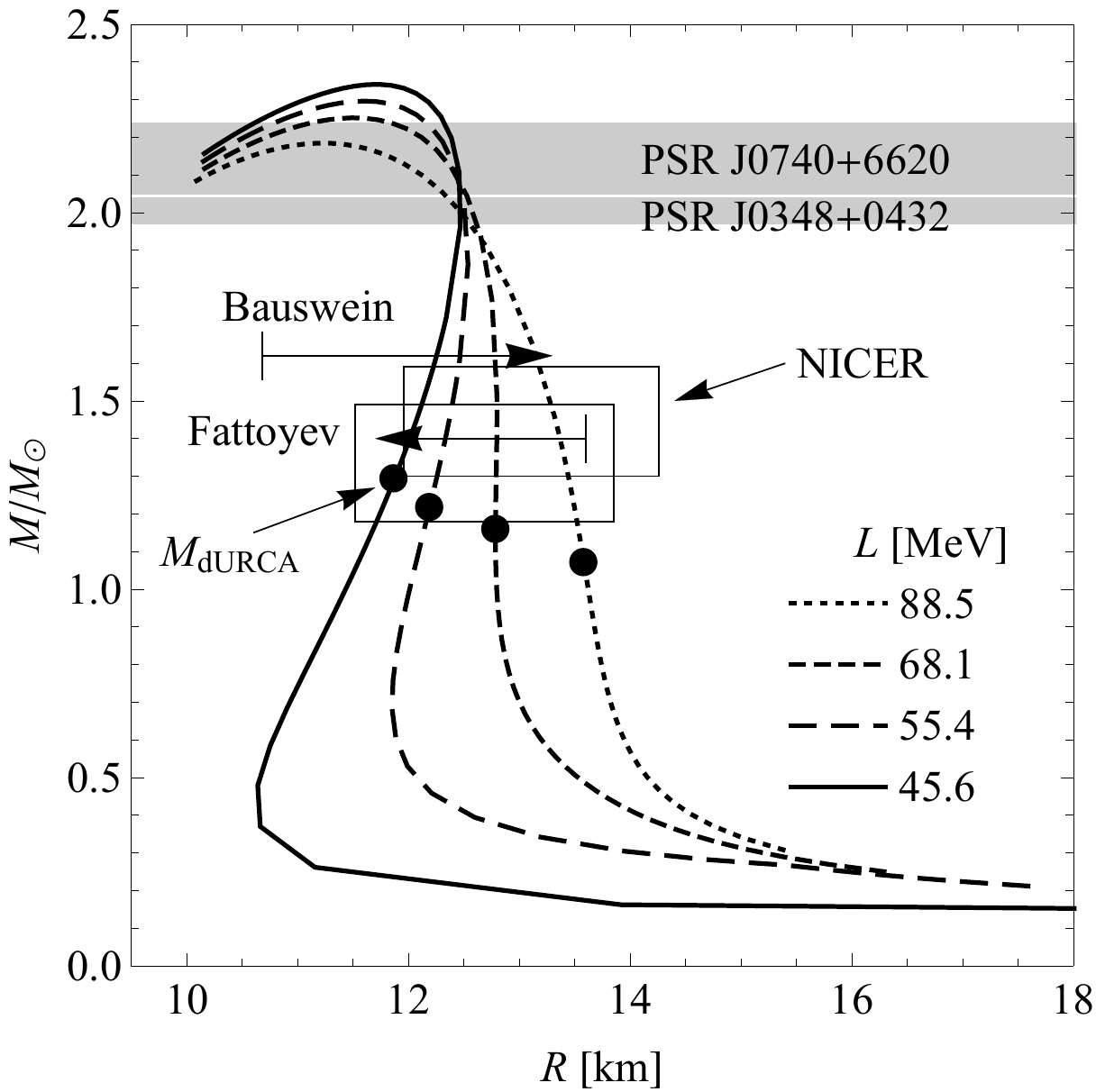}
\caption{Mass-radius profiles for various $C_\delta^2$ corresponding to different $L$. The gray bands indicate
 the most massive stars. The horizontal arrows show the upper and lower bounds for stellar radius from GW180718 analysis.
 The rectangles show the NICER results.}
\label{MvsR}
\end{center}
\end{figure}

The fundamental relationship between stellar parameters is the mass-radius relationship. The $M\!-\!R$ relationship for various
 $C_\delta$ couplings is presented in Fig.~\ref{MvsR}. 

For all considered couplings, the maximum mass is well above $2 ~\rm M_\odot$,
which is consistent with most recent observational data
 \cite{Demorest:2010bx,Antoniadis:2013pzd,Linares:2018ppq}.
In particular, the models with $L$ below 60 MeV are in the best agreement with the most massive pulsar, PSR J0740+6620. 
Promising results concerning the NS radii were obtained from 
the GW signal coming from the binary NS merger \cite{TheLIGOScientific:2017qsa}.
These results suggest a rather small radius for the star, which indicates soft EOS. However, typical soft EOSs lead to $M_{max}$ lower than those for the most massive PSR. A specific EOS is required to find a balance between these two contradictory facts. It should have different behaviors in different ranges of density. The stiffness of the EOS at higher densities controls the maximum mass of a star, whereas at lower densities (closer to the crust) it refers to the stellar radius. 
{ The inclusion of the $\delta$ meson itself (without any crossing terms) makes the EOS slightly stiffer in the broad range of densities and moves both maximum mass and star radii to higher values, as was shown in 
Refs.~\cite{Kubis:1998jt,Menezes:2004vr}.}
The results of this work clearly show that the models with scalar-scalar meson interactions make the EOS softer at moderate densities and stiffer at higher densities. This type of EOS comes from a specific form of the symmetry energy being very soft at a lower-density region. This effect is most  visible for the quadratic coupling model with $L$ 
around 50 MeV. Fattoyev {\em et al.}
 \cite{Fattoyev:2017jql} have already shown that soft symmetry energy is favorable regarding the results of GW170817.
This was concluded from the analysis of the tidal deformability of binary components during the merging. Tidal 
deformability is sensitive to the compactness ($M/R$) of a star, and thus it is possible to find the upper bound of the radius 
for a given mass.
 Another constraint on stellar radius comes from the threshold mass for the prompt collapse into a black hole after the merging. Bauswein {\em et al.} \cite{Bauswein:2017vtn} inferred that the radius of a star with $M=1.6M_\odot$ 
cannot be smaller than 10.7 km. 
Both upper and lower bounds for the radius are shown in the Fig.~\ref{MvsR}. As one may see, the EOS for all 
$L$ are placed within the allowable region. However, the models with $L<60 ~\rm MeV$ have the $M_{max}$ in better agreement with the present constraints from PSR mass measurement.
The mass-radius relation for the EOSs with phase transition shows characteristic bending. For masses smaller than
 $0.5 M_\odot$, the stellar radius decreases with mass. For such low masses, the central density is located in the two-phase region where the EOS is very soft. When the stellar mass is greater, the central density enters into a homogeneous matter region where the EOS becomes much stiffer. Gravitation is no longer able to compress the matter in the core, and the further increase of the star mass causes an increase of the star radius.
At the end of the discussion concerning the mass-radius relation, we should mention the most recent results obtained by the NICER X-ray observatory. The Bayesian analysis of spectral-timing data of PSRJ0030+0451 allowed, for the first time, 
for the simultaneous measurement of the stellar mass and radius. The estimations of the NS star radius and mass from the 
NICER analysis performed by \cite{Miller2019} and \cite{Riley2019} are indicated in Fig.~\ref{MvsR}.

In this paper, we also show the influence of  $\sigma$-$\delta$ meson interactions on the Urca process in $npe\mu$ matter. The direct Urca (Durca) process plays a crucial role in the cooling history of a neutron star.
Applying the Durca threshold given by Eq. (\ref{urca})
\begin{equation}
\begin{split}
&\left( x - \left( (1-x)^{1/3} - x^{1/3} \right)^3 \right)^{2/3} =\\
 &\left( (1-x)^{1/3} -x^{1/3} \right)^2 - \left( \frac{m_\mu}{(3 \pi^2n)^{1/3}} \right)^2~,
\end{split}
\label{urca}
\end{equation} 
the $x^{urca}$ and corresponding $n^{urca}$ are found. In Fig.~\ref{MvsR}, the mass of a star with central density exceeding the threshold $n^{urca}$ is indicated by black points.
A recent analysis of the cooling history of transient system MXB 1659-29 \cite{Brown:2017gxd} strongly suggests 
the presence of rapid cooling of a neutron star core via Durca. 
Direct Urca is only allowed when the symmetry energy increases with density at a sufficiently rapid rate. In our model 
family, the symmetry energy dependence shows the opposite trend, as it may even decrease with density 
\cite{Zabari:2019ukk}. However, this 
behavior changes at some point, and again, the symmetry energy rapidly increases, making the proton fraction sufficiently 
large to ensure Durca, as seen in Fig.~\ref{xplots}. That is why the critical stellar masses for Durca (Fig.~\ref{MvsR}) are still within 
the typical range of neutron star masses.
\begin{figure}[t!]
\begin{center}
\includegraphics[width=\columnwidth]{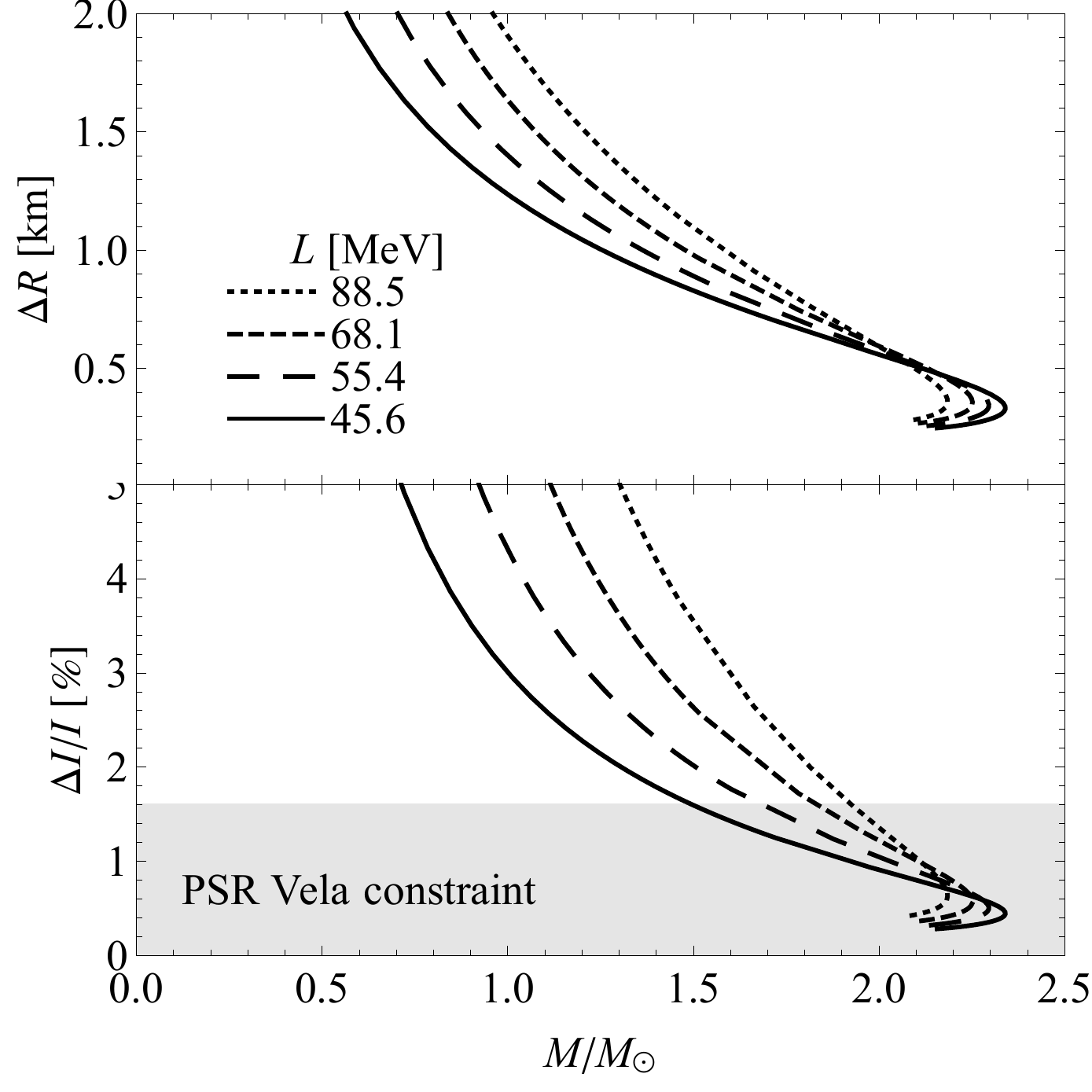}
\caption{The crust thickness and crustal moment of inertia as a function of stellar mass for different nuclear models.}
\label{IvsM}
\end{center}
\end{figure}

Two other parameters describe neutron star properties associated with their crust: the crust thickness
 $\Delta R$ and the crustal fraction of the total moment of inertia $\Delta I/I$. These parameters have particular significance due to the possibility of their reliable observational evaluation from thermal relaxation of the crust and pulsar glitches \cite{Lattimer:2006xb}. 
A glitch phenomenon is a sudden increase in rotational frequencies that many pulsars exhibit. The analysis of these two parameters, based on the 
average rate of angular momentum transfer between the superfluid component and the crust \cite{Link:1999ca}
 allows for estimating a star’s crustal moment of inertia compared to its total moment of inertia.
Thus, the glitching phenomenon confirms that the crust-core transition is of particular importance. The more recent study of cumulative angular momentum transfer of the Vela Pulsar
 \cite{Andersson:2012iu} points out that crust 
should encompass more than $1.6 \%$ of the total moment of inertia. 
{In the approximation of the slow rotation, the total moment of inertia $I$ and its crustal fraction $\Delta I$ may be expressed by the properties of nonrotating configuration described by the TOV equations, as was extensively discussed by
Ravenhall and Pethick \cite{ravenhall1994}. Here we applied the formulas numbered (11) and (16) in their work.}
The results are shown in Figure \ref{IvsM}.
\begin{table}
  \caption{Basic neutron star properties for the quadratic model family}
\label{tab-ns}
\begin{tabular}{|p{2cm}|p{1cm}|p{1cm}|p{1cm}|p{1cm}|}
\hline
 $C_{\delta }^2~\rm [fm^{2}]$ & 1.0 & 3.0 & 3.5 & 3.8 \\
 \hline
 $L~\text{[MeV]}$ & 88.5 & 68.1 & 55.4 & 45.6 \\
 \hline\hline
 $M_{\max }/M_\odot$ & 2.18 & 2.25 & 2.27 & 2.34 \\
 \hline
 $R_{\max }~\rm [km]$ & 11.23 & 11.51 & 11.61 & 11.71 \\
 \hline
 $n_c^{\max }~\rm [fm^{-3}]$ & 1.001 & 0.932 & 0.896 & 0.865 \\
 \hline\hline
 $n_{\text{urca}}~\rm [fm^{-3}]$ & 0.315 & 0.358 & 0.364 & 0.360 \\
 \hline
 $M_{\text{urca}}/M_\odot$ & 1.07 & 1.16 & 1.22 & 1.29 \\
 \hline\hline
 $n_{c-c}$ & 0.0700 & 0.0648 & 0.0608 & 0.0577\\
 \hline
\end{tabular}
\end{table}
A promising method for simultaneous measurement of neutron star mass and radius are based on observations of the 
quiescent mode in the low-mass x-ray binaries (LMXBs) \cite{Ozel:2008kb}.
Precise measurement of $M$ and $R$ is still out of reach, as the uncertainties for both of these quantities are quite 
significant. However, combined results from many observed LMXBs show an interesting tendency: that observed stellar radius 
increases with mass \cite{Ozel:2016oaf}.
This type of relationship between radius and mass is typical for the EOS, being soft in the low-density region and becoming stiffer 
at higher densities. Such an EOS is achieved in our model for low values of the symmetry energy slope  $L$ (see Fig.~\ref{MvsR}).

\section{SUMMARY}

In this work, we showed the neutron star properties obtained in the framework of the recently proposed RMF model of nuclear interactions where the scalar mesons crossing term was applied \cite{Zabari:2019ukk}. Thanks to this new type of coupling, the symmetry energy slope $L$ may reach sufficiently low values, as suggested by the results of terrestrial experiments concerning heavy-ion collisions. Simultaneously, the model leads to the neutron stars which properties are in agreement with the most recent observational data concerning the NS masses and radii. Although less persuasive, the cooling data suggest that fast cooling by direct Urca cycle is present in neutron stars. Typically, low values of symmetry energy block the Durca cycle by the low proton abundance in the NS core. It therefore calls into question the presence of fast cooling. However, the proposed model has interesting properties that, despite the low symmetry energy slope, can still achieve a proton fraction sufficiently large to ensure the presence of the Durca for stars with typical masses.

An additional outcome of the scalar meson couplings leads to phase separation in the outer part of the NS core. One may expect that nuclear matter with separated phases will acquire solid-state properties. The presence of such secondary crust in the NS interior would have interesting implications requiring further analysis.

\bibliography{delta-sigma-stars.bbl}

\end{document}